\documentclass[twocolumn]{aastex62}

\usepackage{lineno}

\graphicspath{./}

\begin{document}

\title{New pulsating stars detected in EA-type eclipsing binary systems based on TESS data}

\correspondingauthor{Xiang-dong Shi}
\email{sxd@ynao.ac.cn}

\author{Xiang-dong Shi}
\affiliation{Yunnan Observatories, Chinese Academy of Sciences(CAS), P.O. Box 110, 650216 Kunming, P.R. China}
\affiliation{University of Chinese Academy of Sciences, Yuquan Road 19\#, Sijingshang Block, 100049 Beijing, China}

\author{Sheng-bang Qian}
\affiliation{Yunnan Observatories, Chinese Academy of Sciences(CAS), P.O. Box 110, 650216 Kunming, P.R. China}
\affiliation{University of Chinese Academy of Sciences, Yuquan Road 19\#, Sijingshang Block, 100049 Beijing, China}

\author{Lin-Jia Li}
\affiliation{Yunnan Observatories, Chinese Academy of Sciences(CAS), P.O. Box 110, 650216 Kunming, P.R. China}

\begin{abstract}
Pulsating stars in eclipsing binaries are very important to understand stellar interior structures through astroseismology because their absolute parameters such as the masses and radii can be determined in high precision based on photometric and spectroscopic data. The high-precision and continuously time-series photometric data of the Transiting Exoplanet Survey Satellite (TESS) provides an unprecedented opportunity to search for and study this kind of variable stars in the whole sky. About 1626 Algol type (EA-type) eclipsing binary systems were observed by TESS in the 1-45 sectors with 2-minutes short cadence. By analyzing those TESS data, we found 57 new pulsating stars in EA-type binary stars. The preliminary results show that those binary systems have orbital periods in the range from 0.4 to 27 days, while the periods of pulsating components are in the range from 0.02 to 5 days. It is detected that 43 targets follow the correlation between pulsation and orbital periods of oscillating eclipsing binaries of Algol type (oEA), which may indicate that they are typical oEA stars. The other 14 targets may be other types of variable stars in eclipsing binary systems. These objects are a very interesting source to investigate the binary structures and evolutions as well as to understand the influences of tidal forces and mass transfer on stellar pulsations.
\end{abstract}


\section{Introduction} \label{sec:intro}

Binary systems are the most reliable object to determine absolute stellar parameters (mass, radius, luminosity, etc.) based on photometric and spectroscopic observations, which is very important to establish stellar evolutionary models. Generally, the internal structure cannot be observed directly for stars, but it can be inversed by astroseismology for pulsating stars. Therefore, binary systems containing at least one pulsating component have greatly attracted the interest of researchers. In addition to the physical parameters obtained by the analysis method of binary systems, the interior structure of the pulsating component can also be analyzed by the method of astroseismology, which is extremely useful for us to understand the stellar structure and evolution, and the tidal interactions on stellar pulsations (e.g. \citet{2017MNRAS.472.1538F, 2020MNRAS.498.5730F, 2021FrASS...8...67G}).

In the early 1970s, some researchers \citep{1971IBVS..596....1T, 1974A&A....34...89B, 1977IBVS.1334....1M} have observed the existence of eclipsing binaries with a pulsating component. However, very few such objects were discovered until the last 20 years after the observation of space telescopes such as CoRot and Kepler (e.g. \citet{2010Ap&SS.328...91D, 2011MNRAS.414.2413S, 2013A&A...552A..60M, 2019A&A...630A.106G}) and other telescopes on the ground (e.g. \citet{2012MNRAS.422.1250L}). So far, the number of such targets has reached hundreds or more (e.g. \citet{2010arXiv1002.2729Z, 2017MNRAS.465.1181L}, etc.), among which a large part of them (more than 200) are $\delta$ Sct component in the classical semi-detached Algols (e.g. \citet{2006MNRAS.370.2013S, 2010arXiv1002.2729Z, 2012MNRAS.422.1250L, 2017MNRAS.465.1181L, 2017MNRAS.470..915K}). This kind of star is defined as oEA (oscillating eclipsing binaries of Algol type) by \citet{2002ASPC..259...96M}. Of course, the EA-type eclipsing binaries can be semi-detached or detached systems. And some pulsating components in EA-type eclipsing binaries are not the primary star or not in semi-detached systems, which are different from the original definition of oEA systems.

\begin{figure*}\centering \vbox to9.0in{\rule{0pt}{5.0in}}
\includegraphics{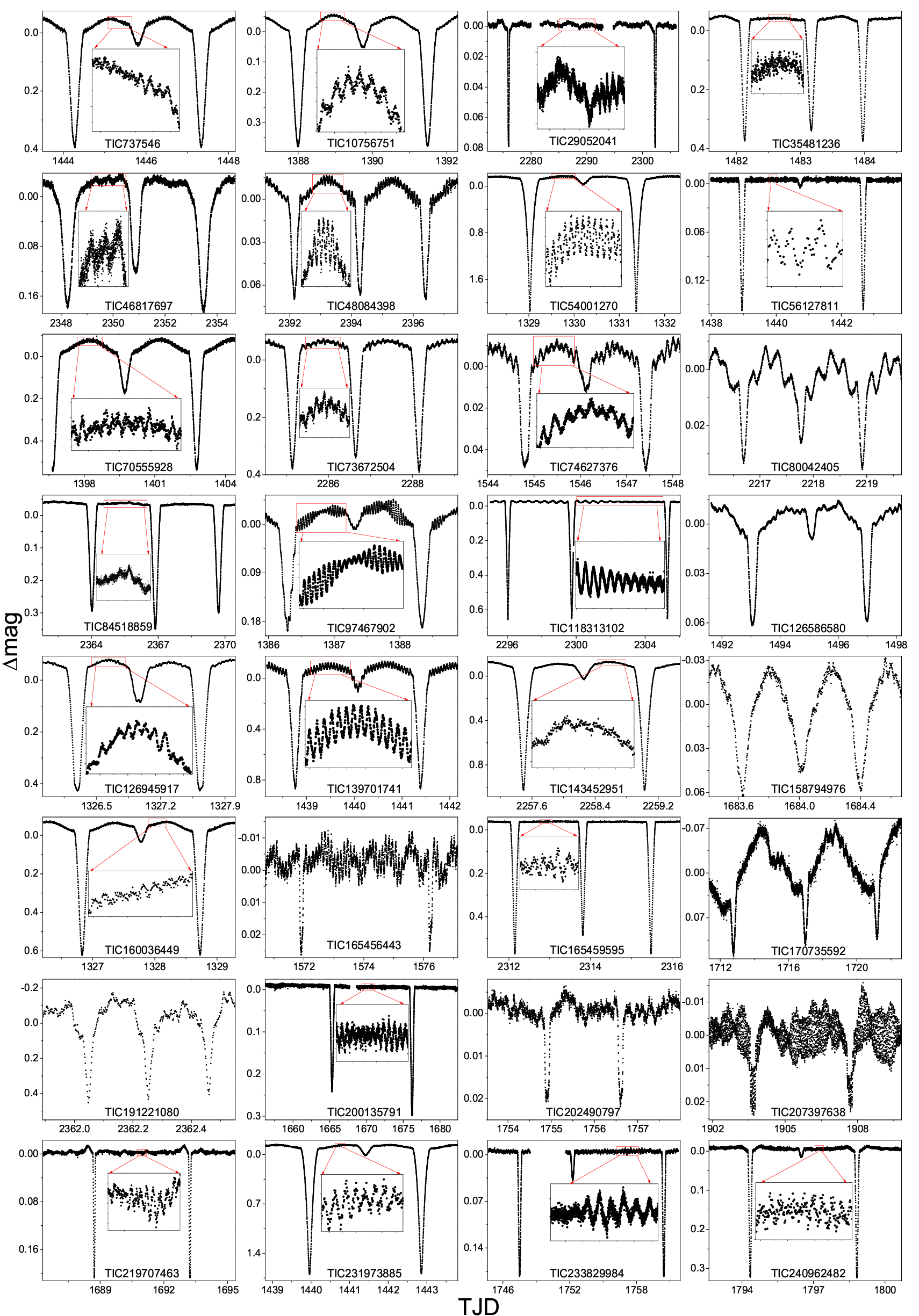}
\caption{The light curves of new eclipsing binary systems with pulsating components (TJD=BJD-2457000.0). TESS ID of each target is marked in each subgraph. Some targets have obvious pulsating variation, but others only can be seen in an enlarged view.}
\label{fig:1}
\end{figure*}

Generally, the mass transfer from the secondary to primary (e.g. \citet{2017ApJ...837..114G, 2021MNRAS.505.3206M}) makes the primary component of oEA stars inside the bottom of the classical Cepheid instability strip and showing $\delta$ Sct-like oscillations with a frequency range of 3-80 cycles $d^{-1}$. Thus, the pulsation of the primary component should be associated with the existence of the companion star . \citet{2006MNRAS.366.1289S} proposed for the first time that there is a correlation between the orbital ($P_{orb}$) and dominant pulsation ($P_{pul}$) periods for oEA stars, while \citet{2006MNRAS.370.2013S} published a catalog including 25 oEA stars. \citet{2012MNRAS.422.1250L} published a catalogue including 74 oEA stars and updated the correlation between $P_{orb}$ and $P_{pul}$. \citet{2013ApJ...777...77Z} made a first theoretical explanation for this relationship. \citet{2017MNRAS.465.1181L} published a catalog including 199 oEA stars.

In April 2018, the Transiting Exoplanet Survey Satellite (TESS, \citet{2015JATIS...1a4003R}) was launched by NASA as an exoplanet  survey mission. It monitored bright stars with a 2-min short cadence and provided full-frame images every 30 minutes. The primary TESS mission is to search for planets transiting bright and nearby stars in all-sky. Meanwhile, its high-precision photometric data also provides unprecedented opportunities for the study of binaries and variable stars.

Due to the observational characteristics of the pulsating stars in eclipsing binaries, the high-precision and continuous time-series photometry data of space telescope have incomparable advantages for the discovery and research of oEA stars. These observational features include that the low-amplitude pulsations of the component are contained in the light curve of binary stars, and their orbital periods are usually greater than or close to 1 day, and so on. In this paper, we report the detection of new pulsating stars in EA-type eclipsing binaries of TESS and make a preliminary analysis of the pulsation properties of these objects.

\section{The pulsating stars in EA-type eclipsing binaries of TESS} \label{}

In the catalog of VSX (the international variable star index, \citet{2006SASS...25...47W}), 95914 EA-type eclipsing binary systems were listed by 2022 January 15. We searched these targets in the TESS catalog of 2-min short cadence from 1-45 sectors based on the criterion $Dist < 5^{"}$, where $Dist$ is the distances (in arcsec) between the two positions determined by the coordinates given in TESS and VSX. It was found that 1626 EA-type binary eclipsing systems were observed by TESS. We have also carried on cross-match these sample with the catalog of 4584 eclipsing binaries observed by TESS \citep{2022ApJS..258...16P}, and a total of 889 objects are listed in this catalog. The information of the first 50 EA-type binary eclipsing systems are listed in Table \ref{tab:EA} and a total of 1626 objects are given as Supporting Information with the electronic version of the paper. In this table, these objects are listed in the order of increasing TESS ID, and these parameters of Column 2 to Column 7 are from the VSX catalog.

\begin{figure*}\centering \vbox to9.0in{\rule{0pt}{5.0in}}
\includegraphics{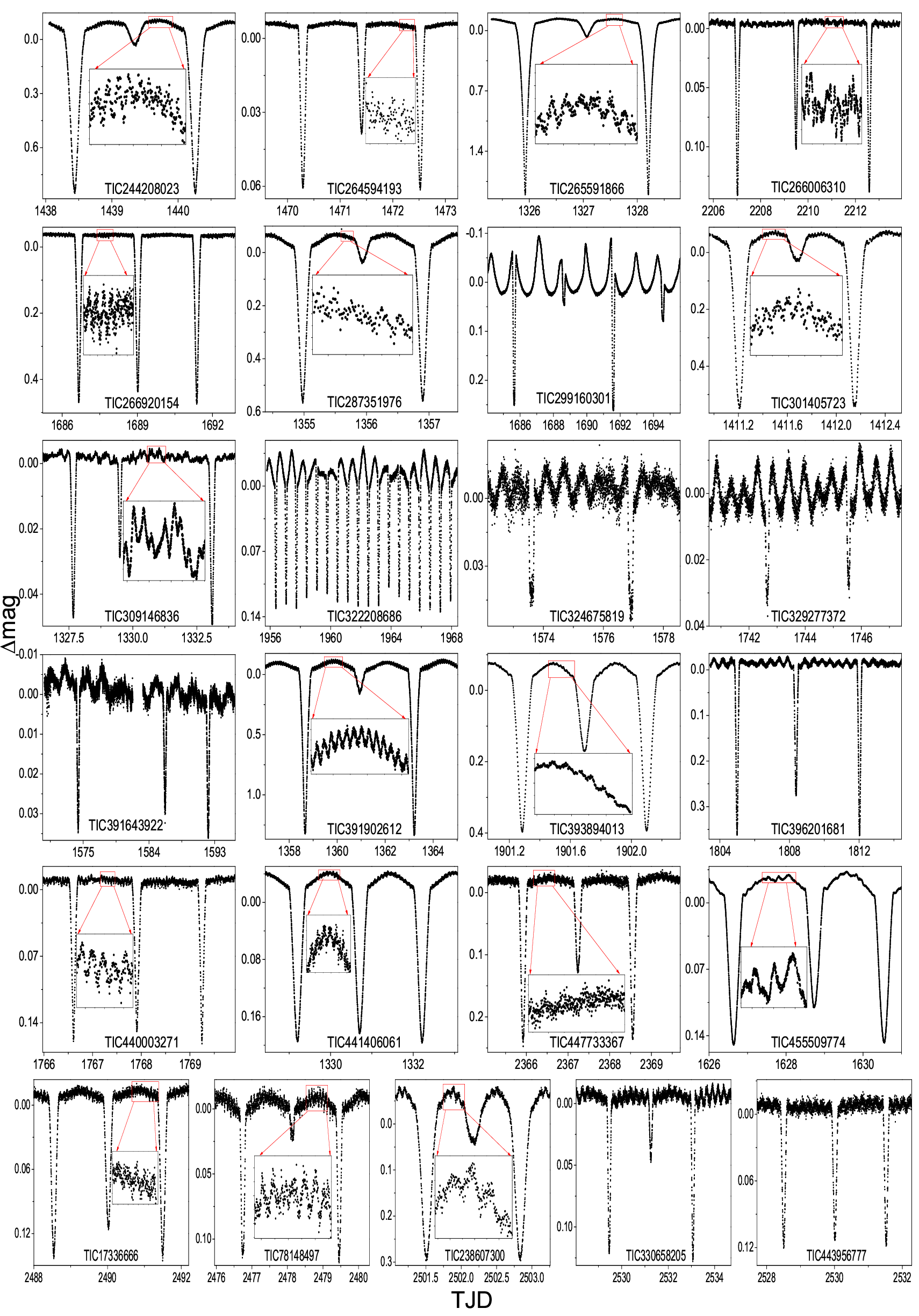}
\caption{Same as those shown in Figure \ref{fig:1} but for different targets.}
\label{fig:2}
\end{figure*}

We downloaded the simple aperture photometry data of these 1626 EA-type binary systems from the Mikulski Space Telescope Archive (MAST) database, and used the steps described by \citet{2021AJ....161...46S, 2021MNRAS.505.6166S} to process the original light curves. Then we examined the light curve of each target with a criterion whether there are obvious regular variations in the light curves outside the eclipse, and searched the published literature. A total of 57 new eclipsing binary systems with possible pulsating components were detected. Their light curves are displayed in Fig. \ref{fig:1} and Fig. \ref{fig:2}, where TJD equals BJD minus 2457000.0. As shown in these two figures, there are similar pulsating variation in the light curve of each target. Meanwhile, some targets have obvious pulsating variation, but others only can be seen in an enlarged view. The information of these objects with the pulsating component are listed in Table \ref{tab:EAoN} and \ref{tab:EAoU}, while these parameters of Column 2 to Column 4 are from the VSX catalog.

Some EA-type binary systems maybe have a late-type component which are likely to be accompanied by magnetic activity, such as KIC 06669809, KIC 10581918, KIC 10619109, KIC 11175495 \citep{2017A&A...607A..85L}, KIC 6048106 \citep{2016ApJ...833..170L}, and KIC 06852488 \citep{2021AJ....161...46S}. In this case, spots may cover the surface of the late-type component and the light curves of the binary system show O'Connell effect, i.e., the phenomenon that the magnitude of the light maximum near 0.25 phase is different from that near 0.75 phase (e.g. \citet{1969CoKon..65..457M, 1999oaaf.conf..377L, 2014ApJS..212....4Q}). The magnetic activity cycle of stars generally is years or decades, such as the average period of sunspot activity is 11 years. Compared with the orbital period of less than 30 days and observation span of tens days, the spot modulations are shown as harmonic frequency peaks of the orbital frequency in the Fourier spectrum.

\begin{figure*}\centering \vbox to4.0in{\rule{0pt}{5.0in}}
\includegraphics{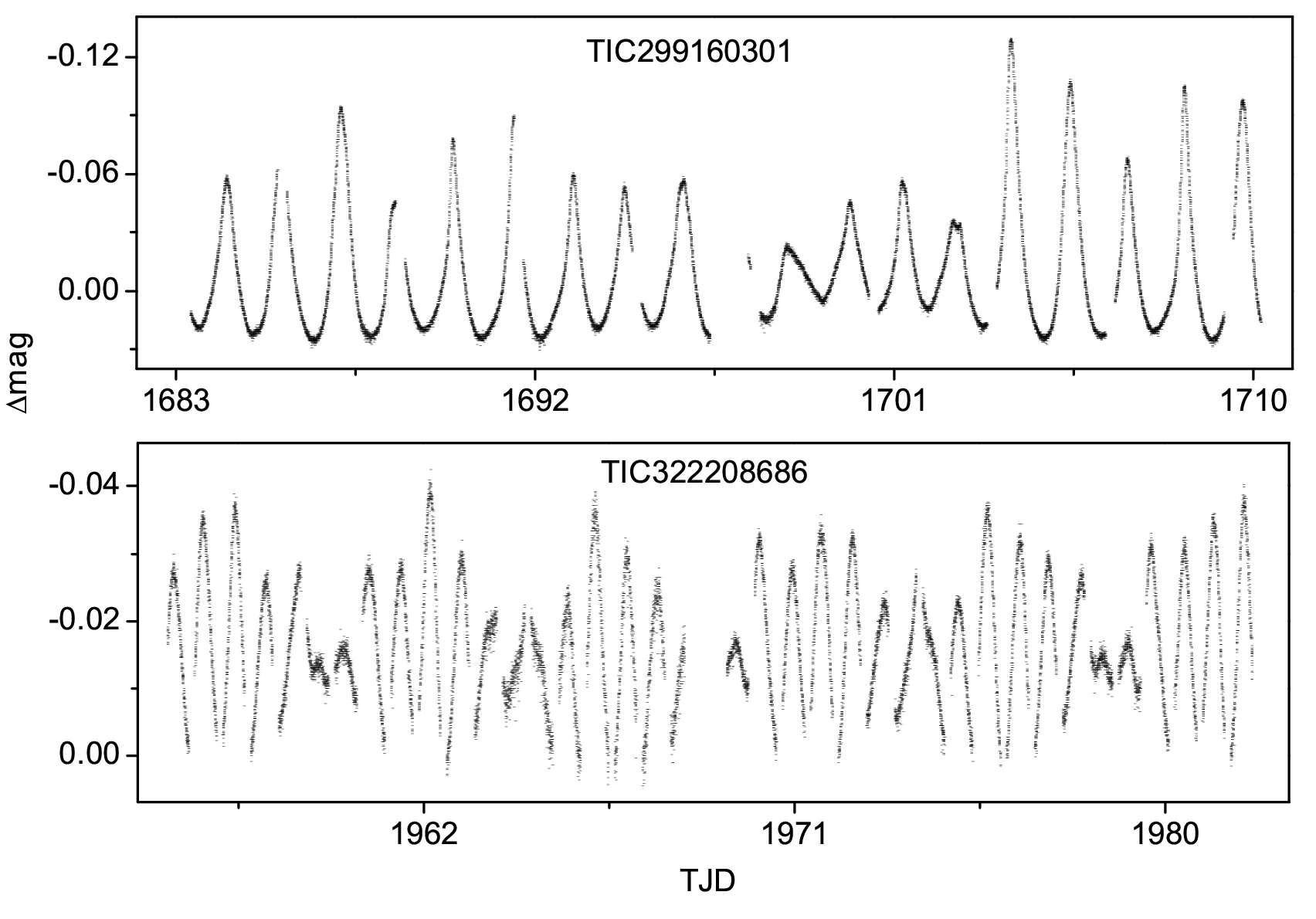}
\caption{The light curves of TIC299160301 and TIC322208686 after removing the data during eclipse.}
\label{fig:299-322}
\end{figure*}

We notice that the light curve of TIC158794976 outside eclipse is continuously changing, which may be $\beta$ Lyrae type (EB-type) or W UMa type (EW-type) eclipsing binary system, i.e., a target misclassified as EA-type eclipsing binary system. At the same time, we also note that the light curves of TIC299160301 and TIC322208686 are different from other targets. After removing the data during eclipse, their curves are shown in Fig. \ref{fig:299-322}, which are similar to typical multi-period pulsating stars and may be $\gamma$ Dor in EA-type eclipsing binary system.


\clearpage

\begin{table*}[]
\begin{center}
\caption{The catalog of EA-type eclipsing binaries observed by TESS (the first 50 lines of the whole table).}
\label{tab:EA}
\begin{tabular}{lllllllllllllll}
 \hline
$TESSID$  & $Name$      &  $Coords(J2000)$ &  $Const.$ & $Period(d)$ & $Mag.range$ & $Dist(")$\\
\hline
TIC737546   & ASAS J045804-2956.0          & 04 58 03.61 -29 55 58.8 & Cae & 3.06772     & 10.84 - 11.32 V   & 0.060 \\
TIC862763   & V0636 Hya                    & 09 01 13.93 -02 23 22.0 & Hya & 2.031716    & 11.83 - 12.55 V   & 0.062 \\
TIC927554   & V0426 Hya                    & 08 28 35.23 -13 51 14.0 & Hya & 7.3089      & 11.55 - 12.23 V   & 0.063 \\
TIC968447   & ASAS J215507-0947.9          & 21 55 07.19 -09 47 57.3 & Cap & 1.86357     & 10.298 (0.220) V  & 0.064 \\
TIC1026849  & ASASSN-V J110404.78-000849.8 & 11 04 04.78 -00 08 49.8 & Leo & 1.5507      & 11.75 - 11.87 V   & 0.015 \\
TIC1045298  & ASAS J021742-0816.7          & 02 17 42.41 -08 16 39.2 & Cet & 1.46344     & 11.16 (0.33) V    & 0.100 \\
TIC1230647  & TIC 1230647                  & 09 09 23.06 -15 35 20.5 & Hya & 1.19        & 9.5 (0.032)TESS V & 0.035 \\
TIC1450518  & ASAS J050049-3212.7          & 05 00 48.98 -32 12 39.7 & Cae & 0.333747    & 12.38 - 12.97 V   & 0.036 \\
TIC1538794  & NSVS 15865206                & 11 07 10.64 -14 38 16.7 & Crt & 0.411427    & 13.25 (0.47) CV   & 0.041 \\
TIC2429093  & ASASSN-V J051947.03+321307.8 & 05 19 47.03 +32 13 07.9 & Aur & 3.7703      & 11.78 - 11.89 V   & 0.067 \\
TIC2761545  & ASASSN-V J234626.75-122918.9 & 23 46 26.75 -12 29 18.9 & Aqr & 6.2066      & 11.93 - 12.25 V   & 0.047 \\
TIC3816260  & ASAS J004032-0628.9          & 00 40 32.41 -06 28 51.2 & Cet & 2.62571     & 10.59 (0.13) V    & 0.095 \\
TIC4247738  & CSS\_J074118.8+311434         & 07 41 18.81 +31 14 34.2 & Gem & 1.30224     & 15.80 (0.25) CV   & 0.455 \\
TIC4742129  & ASAS J023909-1027.8          & 02 39 08.75 -10 27 46.5 & Cet & 1.03598     & 12.34 (0.63) V    & 0.072 \\
TIC4783276  & ASASSN-V J071829.00-201900.0 & 07 18 29.00 -20 19 00.0 & CMa & 2.1689      & 12.35 - 12.54 V   & 0.060 \\
TIC5001059  & ASASSN-V J070946.79-011513.7 & 07 09 46.79 -01 15 13.8 & Mon & 23.129      & 11.39 - 11.55 V   & 0.062 \\
TIC5364915  & KELT KS22C001316             & 09 16 50.40 -21 57 16.2 & Hya & 11.14444    & 8.16 (0.022) V    & 0.726 \\
TIC6090723  & ASASSN-V J051923.94+204004.1 & 05 19 23.94 +20 40 04.1 & Tau & 4.1296      & 11.77 - 11.89 V   & 0.045 \\
TIC6400274  & V0362 Tel                    & 18 54 04.94 -51 30 57.6 & Tel & 1.21125     & 9.68 - 9.96 V     & 0.118 \\
TIC6631253  & ASAS J162637-5042.8          & 16 26 37.40 -50 42 49.3 & Nor & 8.87621     & 9.87 - 10.19 V    & 0.044 \\
TIC7849727  & V0572 Lyr                    & 18 21 38.32 +42 10 07.7 & Lyr & 0.9933045   & 10.6 - 11.0 R1    & 0.023 \\
TIC7851729  & V0571 Lyr                    & 18 21 01.96 +44 38 42.7 & Lyr & 1.2525883   & 11.7 - 12.3 R1    & 0.032 \\
TIC8769657  & HAT-225-0003429              & 09 21 28.35 +33 25 58.6 & Lyn & 0.42647     & 14.06 (0.24) CV   & 0.069 \\
TIC8773089  & IQ Cam                       & 04 26 06.87 +54 28 17.5 & Cam & 0.09017945  & 14.48 - 14.63 Rc  & 0.402 \\
TIC9054370  & HD 222891                    & 23 44 38.88 -08 50 55.6 & Aqr & 1.59495     & 7.72 (0.04) HI-1A & 0.353 \\
TIC9146275  & OV Aqr                       & 23 26 08.62 -19 22 23.6 & Aqr & 21.66595    & 8.72 - 8.98 V     & 0.070 \\
TIC9381557  & MQ Eri                       & 04 53 31.51 -06 41 44.9 & Eri & 11.822      & 12.6 - 13.0 V     & 0.045 \\
TIC9391285  & V0390 UMa                    & 11 10 26.95 +36 32 43.8 & UMa & 2.75259     & 11.18 - 11.42 CV  & 0.260 \\
TIC9725627  & WASP-30                      & 23 53 38.03 -10 07 05.1 & Aqr & 4.156736    & 11.9 (0.00700) V  & 0.371 \\
TIC9787257  & ASASSN-V J235710.36-085915.4 & 23 57 10.36 -08 59 15.4 & Cet & 8.278       & 11.33 - 11.45 V   & 0.073 \\
TIC10400181 & NSVS 8061692                 & 17 52 00.41 +30 40 19.7 & Her & 3.4167989   & 11.60 (0.53) R1   & 1.514 \\
TIC10756751 & GP Cet                       & 00 36 55.13 -05 52 26.5 & Cet & 3.4885      & 9.81 - 10.26 V    & 0.025 \\
TIC11119600 & V2239 Cyg                    & 20 15 17.57 +37 31 43.9 & Cyg & 0.61047966  & 11.73 - 12.51 *   & 0.019 \\
TIC11437325 & ASASSN-V J020033.24+600033.0 & 02 00 33.24 +60 00 33.0 & Cas & 1.1467      & 12.57 - 12.90 V   & 0.321 \\
TIC11918748 & V0699 Cyg                    & 20 17 00.33 +39 08 19.6 & Cyg & 1.55152     & 12.0 - 13.0 p     & 0.101 \\
TIC12494582 & V2914 Cyg                    & 20 18 09.46 +38 18 27.9 & Cyg & 1.661885    & 10.2 (0.3:) V     & 0.048 \\
TIC12723924 & EPIC 205919993               & 22 26 58.71 -17 25 28.0 & Aqr & 11.00354925 & 10.14 (0.055) Kp  & 0.018 \\
TIC12790306 & V0453 Cep                    & 22 52 45.80 +60 54 58.6 & Cep & 1.18475     & 7.53 - 7.64 Hp    & 0.319 \\
TIC12825453 & ASASSN-V J021557.21+655706.7 & 02 15 57.21 +65 57 06.7 & Cas & 1.0976      & 14.06 - 14.23 V   & 0.030 \\
TIC12915124 & ASASSN-V J201956.90+375007.3 & 20 19 56.90 +37 50 07.3 & Cyg & 3.164       & 12.00 - 12.25 V   & 0.046 \\
TIC13062255 & ASAS J050205-2842.8          & 05 02 05.46 -28 42 45.7 & Cae & 1.6512397   & 11.049 (0.170) V  & 0.066 \\
TIC13325340 & ASAS J051433-2519.7          & 05 14 32.89 -25 19 41.9 & Lep & 0.89309     & 12.58 (0.5) V     & 0.082 \\
TIC13351941 & V0559 Cas                    & 02 25 40.11 +61 32 58.8 & Cas & 1.58064     & 7.01 - 7.23 V     & 0.040 \\
TIC13623021 & ASASSN-V J085143.68-070744.4 & 08 51 43.68 -07 07 44.4 & Hya & 2.1703      & 11.68 - 11.84 V   & 0.066 \\
TIC14207118 & V2031 Cyg                    & 20 23 51.01 +38 29 34.3 & Cyg & 2.704666    & 8.57 - 8.68 V     & 0.005 \\
TIC14209654 & KELT KC11C048107             & 20 23 35.46 +38 52 56.6 & Cyg & 2.321887    & 8.83 (0.004)Ic V  & 0.021 \\
TIC14226699 & EPIC 210961508               & 03 59 40.83 +22 21 57.5 & Tau & 0.35003422  & 13.56 (0.000) Kp  & 0.103 \\
TIC14307980 & V0648 Hya                    & 09 38 13.49 -01 04 28.2 & Hya & 0.89742     & 12.15 - 12.8 V    & 0.078 \\
TIC14333263 & EPIC 210954046               & 04 03 36.90 +22 14 57.0 & Tau & 0.95030666  & 12.44 (0.005) Kp  & 0.183 \\
TIC14617089 & NSVS 5733998                 & 20 25 15.77 +37 33 17.9 & Cyg & 1.6688988   & 13.35 - 13.83 V   & 0.036 \\
\hline
\end{tabular}
\end{center}
\tablecomments{Table \ref{tab:EA} is published in its entirety in the machine-readable format.}
\end{table*}


\section{The properties of new pulsating star in EA-type eclipsing binaries}

PERIOD04 \citet{2005CoAst.146...53L} is a software package based on classical Fourier analysis, which will be used to analyze the orbital frequency and the dominant pulsation frequency of the light curves of 57 new pulsating stars in EA-type eclipsing binaries. The errors of frequency and amplitude are calculated following \citet{1999DSSN...13...28M}.

We used the PERIOD04 software to analyze the orbital frequency from the light curves of these 57 new targets, and calculate the corresponding orbital periods. These new orbital periods are listed in Column 6 of Table \ref{tab:EAoN} and \ref{tab:EAoU}, where the errors of the orbital periods are derived from the errors of their orbital frequencies using the error transfer formula. Here we did not obtain a reliable orbital period of TIC29052041, TIC200135791, TIC233829984, and TIC391643922 from the light curves with a short time span, because the orbital periods of these four targets are more than 10 days. The new orbital periods of three targets are different from that given by VSX: the new orbital periods of TIC56127811 and TIC191221080 are half of that given by VSX, and the new orbital period of TIC266920154 is twice as long as that given by VSX. As can be seen in Table \ref{tab:EAoN} and \ref{tab:EAoU}, the orbital periods of these 57 targets are distributed in the range of 0.4-27 days.

\begin{figure*}\centering \vbox to4.0in{\rule{0pt}{5.0in}}
\includegraphics{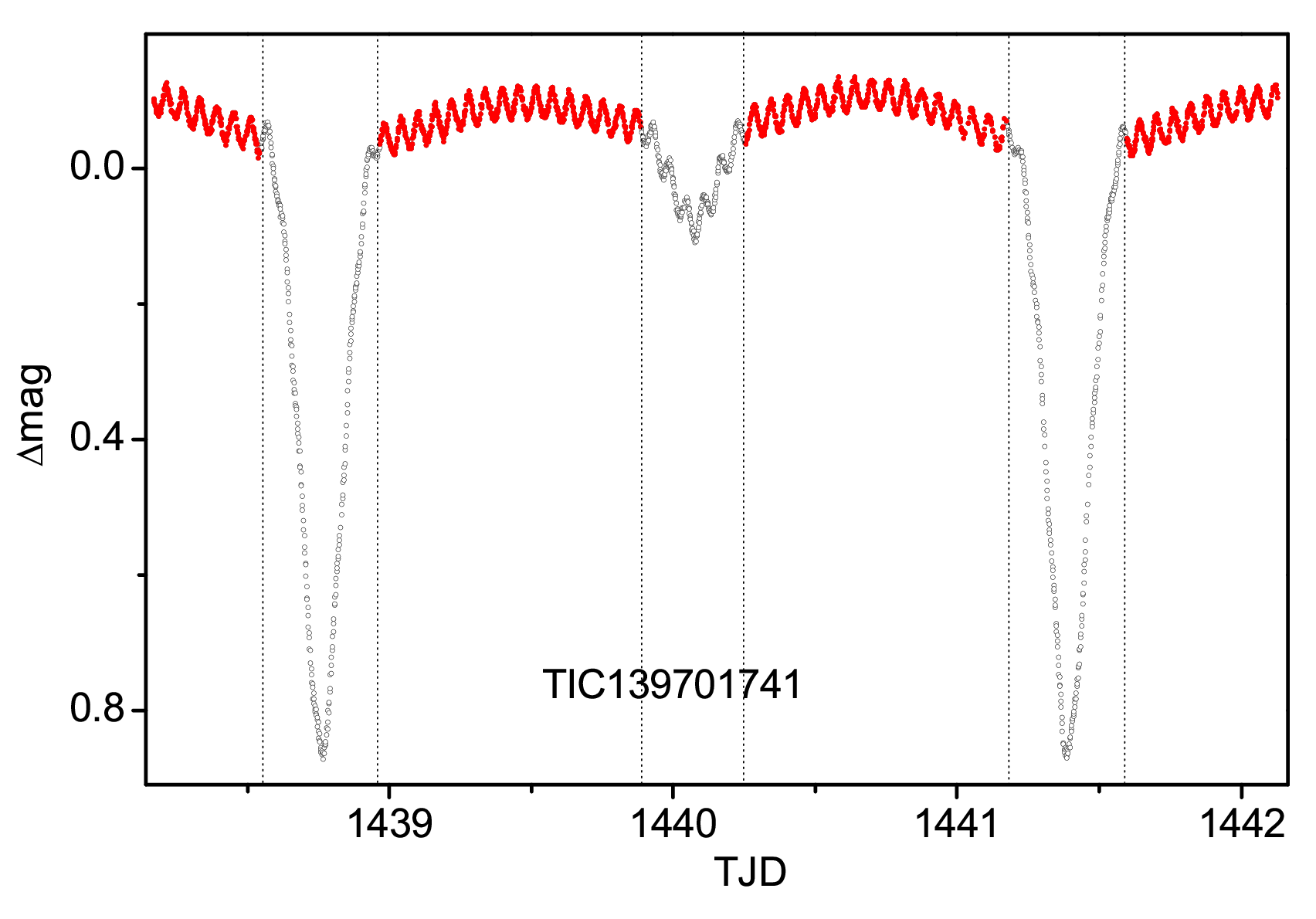}
\caption{Example light curves of TIC139701741. The black open circles refer to the light curve during eclipse. The red solid circles represent the light curve outside eclipse and used to analyze the pulsation frequency.}
\label{fig:139}
\end{figure*}

\begin{figure*}\centering \vbox to4.0in{\rule{0pt}{5.0in}}
\includegraphics{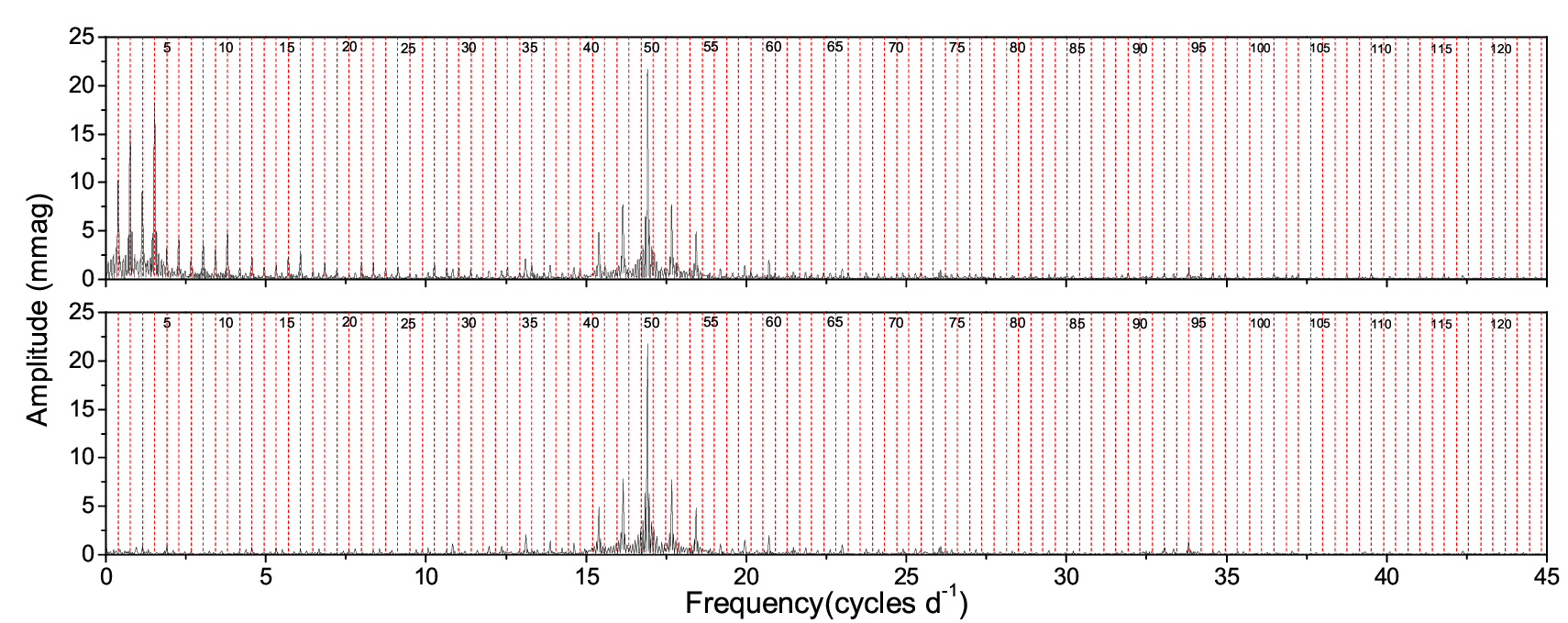}
\caption{Top panel: fourier spectra for the light curve outside eclipse of TIC139701741. Bottom panel: same as top panel, but after pre-whitening all the orbital-harmonic peaks.
The red dotted lines represent the harmonics of the orbital frequency. The most prominent frequency 16.91 cycles $d^{-1}$ is the pulsation frequency from the component.}
\label{fig:139fsp}
\end{figure*}

To analyze the dominant pulsation frequency, we excluded the data during eclipse, such as the light curve of TIC139701741 in Fig. \ref{fig:139}. In this figure, the black open circles and the red solid circles represent the light curve during eclipse and that outside eclipse, respectively. We used the PERIOD04 software to extract and pre-whiten the frequency from the light curve outside eclipse, until obtained a frequency that is not the harmonics of the orbital frequency and can basically fit the most dominant pulsation in the curve. Fourier spectra for the light curve outside eclipse of TIC139701741 are shown in the top panel of Fig. \ref{fig:139fsp}, while the fourier spectra after pre-whitening all the orbital-harmonic peaks are also shown in the bottom panel. In this figure, the red dotted lines represent the harmonics of the orbital frequency, the most prominent pulsation frequency is 16.91 cycles $d^{-1}$, and the second prominent frequency 1.52 cycles $d^{-1}$ is the fourth-order harmonics of the orbital frequency. The most prominent pulsation frequency have sidelobe frequency peaks that are separated by the orbital frequency, and their can be pre-whitened with the prominent pulsation frequency.

The dominant pulsation period and amplitude are listed in Columns 7 and 8 of Table \ref{tab:EAoN} and \ref{tab:EAoU}, where the errors are calculated as described above. As can be seen in these two tables, these 57 targets pulsate in a period range of 0.02-5 days and an amplitude range of 0.09-21 mmag. We could not obtain a pulsation frequency for TIC455509774, which may be due to the aperiodic variation of the light curve. Meanwhile, we also find that the two eclipses of TIC455509774 are not separated by exactly 0.5 in phase, which mean it may has an eccentric orbit.

\begin{figure*}\centering \vbox to5.0in{\rule{0pt}{5.0in}}
\includegraphics{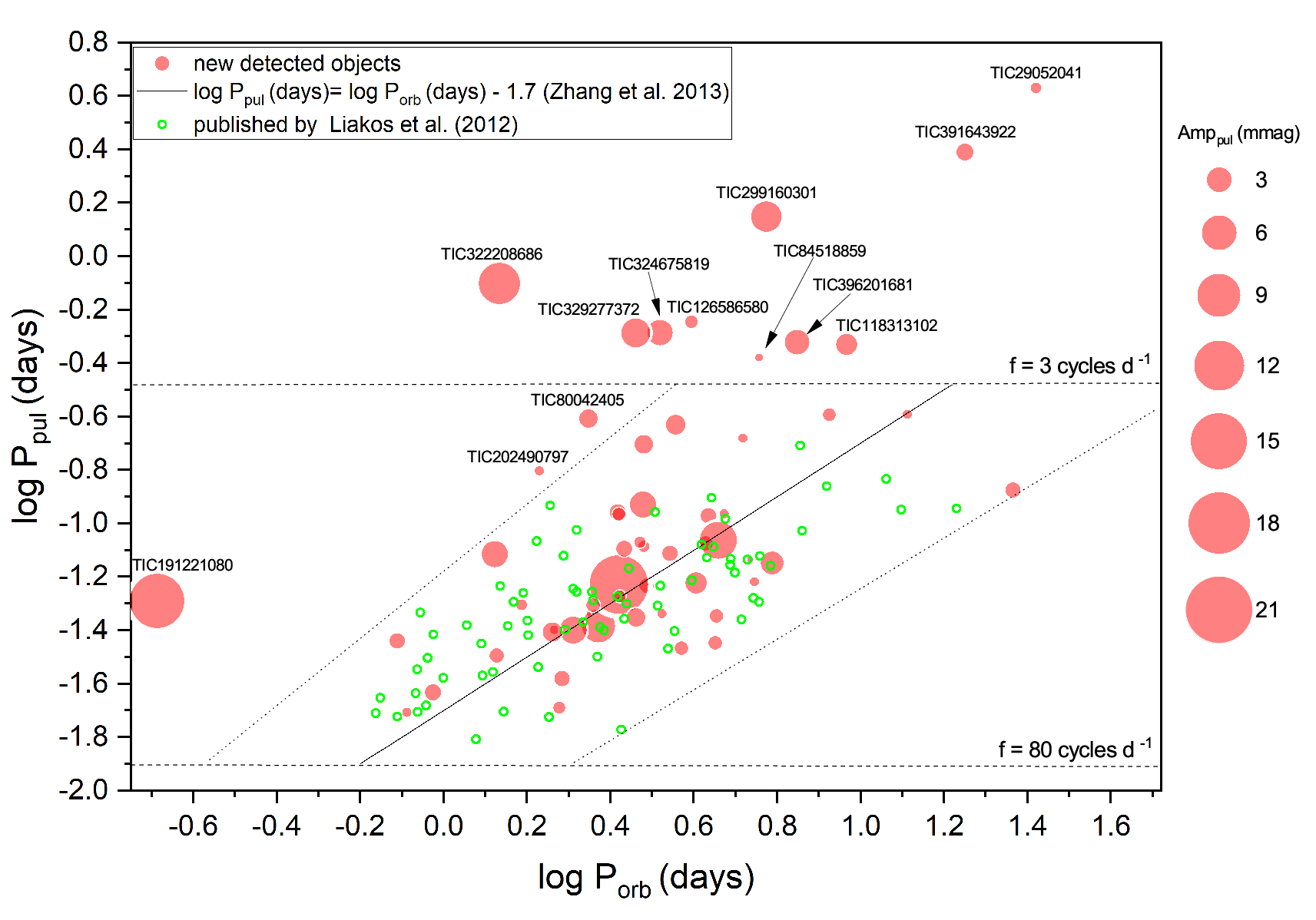}
\caption{The position of new pulsating stars in EA-type eclipsing binaries on the pulsation and orbital-periods relation diagram for oEA stars. The red solid circles refer to these new detected objects, and the size of circles indicate their pulsation amplitude. Meanwhile, these objects (green open circles) published by \citet{2012MNRAS.422.1250L} and the theoretical relation (black solid line) derived by \citet{2013ApJ...777...77Z} are also displayed. }
\label{fig:po-pp}
\end{figure*}

Except for TIC455509774, the position of 56 new pulsating stars in EA-type eclipsing binaries on the $P_{pul}$-$P_{orb}$ relation diagram for oEA stars are shown in Fig. \ref{fig:po-pp} as red solid circles, where the size of circles indicate their pulsation amplitude. The orbital periods of TIC29052041, TIC200135791, TIC233829984, and TIC391643922 are from VSX. In this figure, 43 normal objects show a consistent distribution trend with the theoretical relation (black solid line) derived by \citet{2013ApJ...777...77Z}, where those objects (green open circles) published by \citet{2012MNRAS.422.1250L} are also displayed. As also can be seen in this figure, 13 objects are distinguished from the $P_{orb}$-$P_{pul}$ relation. Among these 13 objects, there are 10 targets pulsating with the frequency of less than 3 cycles $d^{-1}$, and 3 targets pulsating in the frequency range of $\delta$ Sct but deviated from the trend of $P_{pul}$-$P_{orb}$ relation.


\begin{table*}[]
\tiny
\begin{center}
\caption{Catalog of 43 new pulsating stars in EA-type eclipsing binaries following the $P_{pul}$-$P_{orb}$ relation of oEA stars.}
\label{tab:EAoN}
\begin{tabular}{lllllllllllllll}
 \hline
$TESSID$  & $Name$      &  $Coords(J2000)$ &  $Mag.range$ & $Period(d)$ & $P_{Ord}(d)$   & $P_{Pul}(d)$ & $Amp_{Pul}(mmag)$ \\
\hline
TIC737546    &  ASAS J045804-2956.0           & 04 58 03.61 -29 55 58.8 &   10.84 - 11.32 V      & 3.06772     &  3.066414(87)   & 0.0585195(17) &  1.15(3)    \\
TIC10756751  &  GP Cet                        & 00 36 55.13 -05 52 26.5 &   9.81 - 10.26 V       & 3.4885      &  3.49023(17)    & 0.0773914(28) &  1.59(3)    \\
TIC35481236  &  ASAS J064753-1642.9           & 06 47 52.56 -16 42 56.4 &   10.29 (0.4) V        & 1.84343     &  1.84345(35)    & 0.0400270(6)  &  0.48(1)    \\
TIC46817697  &  WISE J152316.6-574419         & 15 23 16.60 -57 44 19.9 &   8.042 (0.367) W1     & 5.1467584   &  5.2209(15)     & 0.208960(38)  &  0.49(4)    \\
TIC48084398  &  V0413 Dra                     & 18 47 29.58 +49 25 55.3 &   7.19 - 7.27 V        & 4.243       &  4.24338(22)    & 0.0841299(15) &  1.14(1)    \\
TIC54001270  &  SW PsA                        & 22 06 09.84 -29 32 54.4 &   10.81 - 13.63 V      & 2.34921     &  2.3503(16)     & 0.0413273(3)  &  7.06(5)    \\
TIC56127811  &  KP Eri                        & 04 43 07.19 -07 24 42.1 &   8.78 - 8.93 V        & 7.447126    &  3.7188(75)     & 0.0341778(3)  &  1.08(1)    \\
TIC70555928  &  AU For                        & 02 15 02.37 -33 51 05.3 &   10.58 - 11.17 V      & 6.11384     &  6.124(11)      & 0.0710720(20) &  3.01(4)    \\
TIC73672504  &  ASAS J102224-3944.7           & 10 22 23.54 -39 44 43.4 &   11.18 (0.43) V       & 3.001494    &  3.0040(18)     & 0.1176729(36) &  4.29(5)    \\
TIC74627376  &  V0483 Vel                     & 09 11 42.32 -46 53 10.4 &   8.23 - 8.30 V        & 2.61362     &  2.6119(13)     & 0.1100297(32) &  1.66(2)    \\
TIC97467902  &  BH Scl                        & 01 34 18.36 -27 21 47.2 &   7.942 - 8.182 Hp     & 2.04507     &  2.0440(14)     & 0.0399145(7)  &  4.42(7)    \\
TIC126945917 &  EE Mic                        & 21 14 55.77 -43 23 43.6 &   9.30 - 9.87 V        & 1.34108     &  1.341079(7)    & 0.0320849(2)  &  1.16(2)    \\
TIC139701741 &  AK Eri                        & 04 25 35.22 -18 48 02.1 &   11.97 - 13.1 V       & 2.63038586  &  2.6315(16)     & 0.0591349(6)  & 21.77(16)   \\
TIC143452951 &  BW Sex                        & 10 21 57.77 -03 43 41.2 &   11.35 - 12.8 V       & 1.537842    &  1.53760(71)    & 0.0495558(32) &  0.66(4)    \\
TIC158794976 &  ASAS J191452+4230.1           & 19 14 52.41 +42 30 08.2 &   10.172 (0.10) V      & 0.772903    &  0.772841(1)    & 0.0364023(6)  &  1.60(4)    \\
TIC160036449 &  RU Gru                        & 22 27 00.55 -37 11 18.1 &   10.98 - 11.78 V      & 1.893193    &  1.89263(74)    & 0.0204061(3)  &  0.88(3)    \\
TIC165456443 &  HD 104186                     & 11 59 51.38 -36 08 53.8 &   9.46 - 9.49 V        & 4.31449     &  4.3145(96)     & 0.1071202(56) &  1.52(3)    \\
TIC165459595 &  V1109 Cen                     & 12 00 46.08 -40 21 16.2 &   9.58 - 10.23 V       & 3.337       &  3.3343(24)     & 0.0460088(13) &  0.55(2)    \\
TIC170735592 &  HAT 199-16913                 & 19 53 54.25 +41 04 47.7 &   11.168 - 11.336 Ic   & 8.412441    &  8.4124(14)     & 0.2557165(42) &  0.95(6)    \\
TIC200135791 &  V0356 Tel                     & 18 37 18.62 -51 54 32.8 &   9.64 - 9.89: V       & 23.1868     &                 & 0.1335113(41) &  1.25(1)    \\
TIC207397638 &  KELT KC22C18110               & 16 06 57.32 +55 26 13.6 &   8.35 (0.009) V       & 4.020534    &  4.0295(53)     & 0.0599633(10) &  2.75(4)    \\
TIC219707463 &  FX UMa                        & 09 06 22.44 +68 26 42.5 &   7.08 - 7.27 V        & 4.507176    &  4.50725(29)    & 0.0450924(1)  &  1.03(1)    \\
TIC231973885 &  TW Pup                        & 06 22 01.29 -47 53 57.1 &   11.28 - 13.8 V       & 2.89481     &  2.8940(25)     & 0.0444477(8)  &  2.13(4)    \\
TIC233829984 &  PMAK V78                      & 20 14 01.01 +58 43 50.2 &   9.21 - 9.39 V        & 12.94875    &                 & 0.2561010(50) &  0.50(2)    \\
TIC240962482 &  V1070 Cas                     & 01 15 58.96 +52 46 40.0 &   10.44 - 10.80 R1     & 4.4745      &  4.475(14)      & 0.0357336(9)  &  1.12(3)    \\
TIC244208023 &  V1637 Ori                     & 04 53 08.59 -03 29 52.8 &   12.10 - 13.22 V      & 1.82276     &  1.82238(79)    & 0.0392377(1)  &  2.32(6)    \\
TIC264594193 &  V1804 Ori                     & 05 23 05.53 +01 03 24.9 &   7.08 - 7.14 V        & 2.22878     &  2.2281(17)     & 0.0459181(29) &  0.09(1)    \\
TIC265591866 &  AQ Ind                        & 22 07 55.05 -59 52 29.6 &   11.04 - 13.75 V      & 2.280821    &  2.2811(13)     & 0.0494380(14) &  1.04(3)    \\
TIC266006310 &  NSV 2932                      & 06 21 25.75 +02 16 06.3 &   6.29 - 6.41 V        & 5.56432     &  5.564(15)      & 0.0604537(18) &  0.49(1)    \\
TIC266920154 &  V2541 Cyg                     & 20 24 11.89 +48 55 26.1 &   9.97 - 10.45: R1     & 2.3542      &  4.7070(16)     & 0.108618(11)  &  0.50(2)    \\
TIC287351976 &  AB Vol                        & 08 10 43.50 -72 32 44.8 &   11.98 - 12.81 V      & 1.92193     &  1.92207(15)    & 0.0262566(1)  &  1.52(3)    \\
TIC301405723 &  KZ Eri                        & 03 31 54.25 -01 38 21.4 &   11.3 - 12.05 V       & 0.94269     &  0.94264(22)    & 0.0233225(3)  &  1.69(5)    \\
TIC309146836 &  HD 69863A                     & 08 15 15.93 -62 54 56.3 &   5.243 - 5.302 V      & 5.42793     &  5.42793(11)    & 0.0732485(1)  &  0.27(1)    \\
TIC391902612 &  AW Men                        & 07 06 16.33 -76 50 21.4 &   12.45 - 14.60 V      & 4.5521      &  4.55206(79)    & 0.0864262(2)  &  8.45(5)    \\
TIC393894013 &  CI CVn                        & 13 13 33.36 +47 47 51.9 &   9.33 - 9.9 Hp        & 0.8158743   &  0.815872(95)   & 0.0196612(3)  &  0.51(2)    \\
TIC440003271 &  V0342 And                     & 00 10 03.19 +46 23 25.1 &   7.58 - 7.72 Hp       & 2.63934     &  2.6387(28)     & 0.0534324(21) &  0.83(3)    \\
TIC441406061 &  BV Mic                        & 20 43 12.48 -32 17 35.4 &   9.84 - 10.09 V       & 3.018225    &  3.0196(14)     & 0.0820754(33) &  0.76(2)    \\
TIC447733367 &  ASASSN-V J170046.29-640051.9  & 17 00 46.29 -64 00 51.9 &   12.27 - 12.58 V      & 2.6312      &  2.6325(16)     & 0.108451(11)  &  0.98(4)    \\
TIC17336666  &  ASASSN-V J042339.45+212019.8  & 04 23 39.45 +21 20 19.8 &   11.74 - 11.89 V      & 2.95544     &  2.9554(22)     & 0.0851887(60) &  0.77(3)    \\
TIC78148497  &  ASASSN-V J055424.47+261831.7  & 05 54 24.47 +26 18 31.7 &   11.15 - 11.24 V      & 2.7099      &  2.70989(49)    & 0.0803482(9)  &  1.42(3)    \\
TIC238607300 &  LL Cnc                        & 08 50 51.20 +19 21 26.2 &   12.02 - 12.48 V      & 1.32434     &  1.324327(93)   & 0.0767742(8)  &  4.17(5)    \\
TIC330658205 &  EPIC 211462458                & 09 05 37.05 +12 35 23.1 &   11.19 (0.118) Kp     & 3.60195759  &  3.6020(48)     & 0.234549(21)  &  2.46(4)    \\
TIC443956777 &  ASASSN-V J081525.20+102352.5  & 08 15 25.20 +10 23 52.5 &   11.94 - 12.12 V      & 3.0203      &  3.0203(35)     & 0.198150(17)  &  2.16(4)    \\
\hline
\end{tabular}
\end{center}
\tablecomments{The one or two digital numbers in the parentheses are the errors on the last one or two bits of the data.}
\end{table*}

\begin{table*}[]
\tiny
\begin{center}
\caption{Catalog of 14 new pulsating stars in EA-type eclipsing binaries not following the $P_{pul}$-$P_{orb}$ relation of oEA stars.}
\label{tab:EAoU}
\begin{tabular}{lllllllllllllll}
 \hline
$TESSID$  & $Name$      &  $Coords(J2000)$ &  $Mag.range$ & $Period(d)$ & $P_{Ord}(d)$   & $P_{Pul}(d)$ & $Amp_{Pul}(mmag)$ \\
\hline
TIC29052041  & KX Vel                     & 08 50 33.46 -46 31 45.1 & 5.08 - 5.16 V     & 26.306047   &                 & 4.2630(15)    &  0.67(1)    \\
TIC80042405  & FU CMa                     & 07 00 19.36 -22 07 08.6 & 6.52 - 6.56 V     & 2.2186      &  2.2227(18)     & 0.247318(46)  &  2.10(3)    \\
TIC84518859  & ASAS J155358-5553.4        & 15 53 57.56 -55 53 21.8 & 9.59 - 10.10 V    & 5.691743    &  5.6998(74)     & 0.41759(13)   &  0.39(1)    \\
TIC118313102 & NT Vel                     & 08 34 24.37 -54 40 03.1 & 8.32 - 9.02 V     & 9.255699    &  9.25576(65)    & 0.4682821(49) &  2.77(1)    \\
TIC126586580 & V0390 Pup                  & 07 44 34.17 -24 40 26.7 & 5.61 - 5.69 V     & 3.9279      &  3.927687(93)   & 0.5681832(36) &  1.00(2)    \\
TIC191221080 & V1596 Sco                  & 16 49 54.47 -29 34 38.8 & 12.7 - 13.5 V     & 0.41056     &  0.205284(5)    & 0.0513214(16) & 18.62(59)   \\
TIC202490797 & KELT KC21C00851            & 15 18 51.28 +63 09 16.4 & 8.44 (0.009) V    & 1.69626     &  1.696176(50)   & 0.1575198(7)  &  0.50(1)    \\
TIC299160301 & V2077 Cyg                  & 19 16 44.50 +50 38 47.9 & 9.16 - 9.31 Hp    & 5.9372      &  5.93720(51)    & 1.41103(26)   &  5.73(10)   \\
TIC322208686 & WISE J214136.6+674539      & 21 41 36.62 +67 45 39.3 & 8.688 (0.130) W1  & 1.3542684   &  1.35964(32)    & 0.792160(40)  & 10.11(3)    \\
TIC324675819 & KELT KS38C021633           & 11 49 05.48 -61 41 32.2 & 10.80 (0.055) V   & 3.302989    &  3.3030(41)     & 0.518573(87)  &  4.14(6)    \\
TIC329277372 & KELT KC24C024065           & 22 10 31.27 +57 16 56.6 & 11.18 (0.022) V   & 2.887904517 &  2.88790(88)    & 0.516980(17)  &  5.08(3)    \\
TIC391643922 & V0736 Car                  & 10 47 38.88 -60 37 04.3 & 7.91 - 8.18 V     & 17.7997     &                 & 2.45539(53)   &  1.79(2)    \\
TIC396201681 & V0961 Cep                  & 23 58 05.99 +67 36 11.4 & 10.43 - 10.83 R1  & 7.03848     &  7.0386(12)     & 0.4779746(19) &  3.92(1)    \\
TIC455509774 & del Cir                    & 15 16 56.90 -60 57 26.1 & 5.04 - 5.20 V     & 3.902476    &  3.9074(22)     &               &             \\
\hline
\end{tabular}
\end{center}
\tablecomments{Same as Table \ref{tab:EAoN} but for different targets.}
\end{table*}

\section{Discussion and conclusion}

Among 95914 EA-type binaries listed in the catalog of VSX, 1626 EA-type binaries are observed by TESS in 2-min short cadence from 1-45 sectors. A total of 57 new pulsating stars are detected from 1626 EA-type binaries, and we analyze their new orbital periods and the dominant pulsation periods. Among the 57 targets, 43 samples show a consistent distribution trend with the $P_{pul}$-$P_{orb}$ relation of oEA stars, which means that they should be oEA stars. Meanwhile, 10 targets pulsate with the frequency of less than 3 cycles $d^{-1}$, which means that they are likely to be $\gamma$ Dor, such as TIC299160301 and TIC 322208686, or other variable stars in EA-type eclipsing binary systems. Three targets (TIC191221080, TIC202490797, and TIC80042405) pulsate in the frequency range of $\delta$ Sct but deviated from the trend of $P_{pul}$-$P_{orb}$ relation, which may mean that they are $\delta$ Sct star in eclipsing binary systems, but maybe with a different formation mechanism from oEA stars.

The light curve of TIC455509774 shows eccentric eclipse and similar pulsating variation, but could not obtain a pulsation frequency, which may be due to aperiodic pulsating variation caused by stellar activities, such as spots or chemical-concentration spots. The light curve of TIC158794976 shows continuous variation outside eclipse, which may be EB or EW-type eclipsing binary system. It is an important object to study the influences of tidal forces on stellar pulsations. It resembles HL Dra \citep{2021MNRAS.505.6166S} where both the primary and secondary components have a high filling factor.

These new objects are very interesting sources for further investigations of binary formation and evolutions and the influences of tidal forces and mass transfer on stellar pulsations. However, many characteristics of these objects are still unclear, which requires more spectral or photometric observations in the future.

\acknowledgments
This work is partly supported by Chinese Natural Science Foundation (Nos. 11933008 and 12103084). The TESS data presented in this paper were obtained from the Mikulski Archive for Space Telescopes (MAST) at the Space Telescope Science Institute (STScI). STScI is operated by the Association of Universities for Research in Astronomy, Inc. Support to MAST for these data is provided by the NASA Office of Space Science. Funding for the TESS mission is provided by the NASA Explorer Program.


\end{document}